\documentclass[aps,reprint,superscriptaddress,nofootinbib]{revtex4-1}

\usepackage[utf8]{inputenc}
\usepackage{amsmath,amsthm,amssymb}
\usepackage{mathtools,dsfont}
\usepackage{braket,bm}
\usepackage[normalem]{ulem}
\usepackage{graphicx,comment}
\usepackage{float,comment}
\usepackage{hyperref,xcolor}
\hypersetup{colorlinks=true,linkcolor=blue,citecolor=blue}

\renewcommand{\comment}[1]{}

\newcommand{\h}{\mathcal{H}}

\newcommand{\Z}{\mathcal{Z}}

\begin{document}
	
	\title{Thermal transitions in a one-dimensional, finite-size Ising model}
	
	\date{\today}
	
	\author{Varazdat Stepanyan}
    \affiliation{A. Alikhanyan National Science Laboratory (YerPhI), 0036 Yerevan, Armenia}
    \affiliation{Physics Department, Yerevan State University,  0025 Yerevan, Armenia}
	\author{Andreas F. Tzortzakakis}
    \affiliation{Institute of Electronic Structure and Laser, Foundation for Research \& Technology –- Hellas, 70013 Heraklion, Crete, Greece}
    \affiliation{Department of Physics, National and Kapodistrian University of Athens, 15784 Athens, Greece}
	\author{David Petrosyan}
	\affiliation{A. Alikhanyan National Science Laboratory (YerPhI), 0036 Yerevan, Armenia}
    \affiliation{Institute of Electronic Structure and Laser, Foundation for Research \& Technology –- Hellas, 70013 Heraklion, Crete, Greece}
    \author{Armen E. Allahverdyan}
	\affiliation{A. Alikhanyan National Science Laboratory (YerPhI), 0036 Yerevan, Armenia}
	
\begin{abstract}
We revisit the one-dimensional ferromagnetic Ising spin-chain with a finite number of spins and periodic boundaries and derive analytically and verify numerically its various stationary and dynamical properties at different temperatures. 
In particular, we determine the probability distributions of magnetization, the number of domain walls, and the corresponding residence times for different chain lengths and magnetic fields. 
While we study finite systems at thermal equilibrium, we identify several temperatures similar to the critical temperatures for first-order phase transitions in the thermodynamic limit.   
We illustrate the utility of our results by their application to structural transitions in biopolymers having non-trivial intermediate equilibrium states.
\end{abstract}

\maketitle

\section{Introduction}

The Ising model for arrays of spin-$1/2$ particles is of fundamental interest in physics and chemistry as it emulates interacting many-body systems \cite{brush}. The model was introduced by Lenz \cite{lenz} and solved exactly by Ising for a one-dimensional (1D) system at equilibrium, where it does not exhibit  thermodynamic phase-transition at finite temperatures \cite{ising}. Ising also provided an ingenious argument for this 1D behavior to generically persist in higher dimensions \cite{ising}, which fortunately turned out to be incorrect \cite{brush}. 
Still, the 1D Ising model attracted much attention as the simplest model where the interactions, noise, size, and dimension of a statistical system are simultaneously important. Moreover, its zero-temperature phase transition is interesting and non-trivial, obeying the hyper-scaling relation \cite{baxter}. An incomplete list of applications of the 1D Ising model includes: simulating quasi-1D systems in (soft) condensed matter \cite{quasi1,quasi2,quasi3}, modeling secondary and tertiary structure in biopolymers  \cite{azbel,zimmo,go_go,book,qian,folding,armen_hc}, and emulating the simplest hidden Markov models -- the standard tool in data science -- in machine learning and probabilistic inference \cite{armen1,armen2}. 
    
Given all the attention to the 1D Ising model for more than 100 years \cite{brush,baxter}, is there anything new to be learned from it? The answer is yes, and we show that for practically relevant situations involving finite numbers of spins, the 1D Ising model exhibits several interesting  temperature-dependent transitions between paramagnetic and ferromagnetic states. While these are not thermodynamic phase transitions, they are still important for describing many relevant physical processes, such as, e.g., secondary and tertiary structural transitions in biopolymers as we demonstrate below. 
These thermal transitions are not revealed by the mean order parameter but are manifest in the probability distributions for magnetization and the number of domain walls, which we derive using an efficient analytical approach and obtain transition temperatures that are similar in several respects to the corresponding critical temperatures in thermodynamic limit. 
We consider canonical ensemble and show in Appendix \ref{microcano} that microcanonical ensemble cannot describe thermal transitions and the related physical effects. The results of the analytic approach are corroborated by the dynamical Monte Carlo simulations that reveal even richer structure of equilibrium states and thermal transitions between them. We note that the equilibrium probability distribution of magnetization was determined in Ref.~\cite{bruce} and also in Refs.~\cite{racz,campo} for various boundary conditions, while analytic results for the distribution of domain walls were presented in Refs.~\cite{quasi2,campo}. Also, the standard transfer-matrix method we employ was recently used in Ref.~\cite{campo} and related to quantum measurement.  To our knowledge, however, the present manuscript presents the first systematic study of thermal transitions in finite-size 1D Ising model.

The paper is organized as follows. After introducing in the next section the Ising model, we present in Sec.~\ref{sec:analytics} an analytical approach for calculating equilibrium probabilities of magnetization and the number of domain walls. 
Here we also determine transition temperatures that are based on the shape of the distribution function for magnetization and we report on new scaling relations for the average size of domains. We verify our results via exact numerical simulations of the dynamics of the spin chain in Sec.~\ref{sec:MCdyn} and determine the residence times for the chain magnetization and domain walls. Here we also define and interpret dynamical transition temperature. In Sec.~\ref{sec:biophys} we use the intuition gained from our analysis to understand the model of helix-coil structural transition commonly employed in biophysics. Our conclusions are summarized in Sec.~\ref{sec:conclusions}. In Appendix \ref{microcano} we presents some results on the equilibrium microcanonical distribution.

\section{Ising spin chain in a thermal environment}
\label{sec:Isingmodel}

Consider a one-dimensional chain of $N$ classical spins $\sigma_j =\pm 1$ described by the Ising Hamiltonian 
\begin{equation}
	\h= -J \sum_{j=1}^N \sigma_j \sigma_{j+1}-h\sum_{j=1}^N \sigma_j, \label{eq:IsingHam}
\end{equation} 
where $J$ is the interaction strength between the neighboring spins, $h$ is the external magnetic field that lifts the degeneracy of $\h$ with respect to single spin-flips $\sigma_j \to \bar{\sigma}_j$, and we assume periodic boundary conditions $\sigma_{N+1}= \sigma_{1}$. 
We consider ferromagnetic interaction $J>0$ that favors the spins aligned in the same direction.
	
A system of $N$ spins has $2^N$ possible microscopic states (spin configurations) $\{ \sigma \} \equiv \{ \sigma_1,\sigma_2, \ldots, \sigma_N \}$. We assume that the system is immersed in a thermal environment at finite temperature $T=1/\beta$ ($k_\mathrm{B}=1$) which induces transitions between these states. 
The probabilities (populations) $p_\mu$ of the spin configurations evolve in time according to the Markov rate equations
\begin{equation}
\dot{p}_\mu=-p_\mu \sum_{\nu} \Gamma_{\mu \nu}+ \sum_{\nu} p_\nu \Gamma_{\nu \mu},
\label{eq:ddtpk}
\end{equation}
where $\Gamma_{\mu \nu} \equiv \Gamma(\mu \to \nu)$ are the transition rates between the microscopic states $\mu$ and $\nu$. We model them using the Glauber rates \cite{Glauber_rates}
\begin{equation}
\Gamma_{\mu \nu}=\dfrac{\Gamma_0}{1+e^{\beta \Delta E_{\mu \nu}}}, \quad \Delta E_{\mu \nu} = E_{\mu}-E_{\nu} , \label{eq:Glauberrates}
\end{equation}
which respect the detailed balance condition 
\begin{equation}
\label{deti}
\Gamma_{\mu \nu} e^{-\beta E_\nu} = \Gamma_{\nu \mu} e^{-\beta E_\mu}, 
\end{equation}
where $E_{\mu,\nu}$ are the energies of configurations $\{ \sigma \}_{\mu,\nu}$. We assume that only single spin-flip transitions are allowed, and the transition rates between configurations that differ by two or more spin-flips vanish, $\Gamma_{\mu \nu}=0$. 
The detailed balance condition (\ref{deti}) ensures that at long times the system attains the Gibbs equilibrium state: 
\begin{equation}
\label{gib}
p_\mu=e^{-\beta E_\mu}/\Z,\quad \Z=\sum_\nu e^{-\beta E_\nu}. 
\end{equation}

\section{Equilibrium analysis of finite 1d Ising chain}
\label{sec:analytics}

\subsection{Analytical approach}
 
Since the configuration space of the system grows exponentially with its size $N$, direct numerical calculations of the microscopic properties of the system through Eq.~(\ref{eq:ddtpk}) become prohibitively difficult already for moderate numbers of spins $N \gtrsim 15$. Yet, for an ergodic system with periodic boundary conditions, the steady-state probability distributions of various macroscopic properties, such as the magnetization and spin correlations, are amenable to analytic treatment leading to expressions that are linear in $N$, as derived below.
	
The Kronecker delta $\delta[n]$ for a discrete variable $n \in [-N, N]$ can be cast as a Fourier series
\begin{equation}
\delta[n] = \frac{1}{2N+1} \sum_{k=-N}^{N}e^{i\lambda_k n}, \quad \lambda_k \equiv \dfrac{2\pi k}{2N+1}.
\end{equation}
Consider any discrete function $f(\sigma) \in [-N,N]$ of the spin configurations $\{ \sigma \}$ with $\sigma_j = \pm 1 \, \forall \, j \in [1,N]$. In the steady state, the probability distribution of various values $n$ of $f(\sigma)$ is then 
\begin{eqnarray}
P(n) &=& \frac{1}{\Z} \sum_{\{\sigma\}} e^{-\beta \h(\sigma)} \delta \big[f(\sigma)-n\big] \nonumber \\
&=& \frac{1}{\Z(2N+1)}\sum_{k=-N}^{N}e^{-i\lambda_k n} \sum_{\{\sigma\}} e^{-\beta \tilde{\h}(\sigma,k)},   \label{eq:P_f}
\end{eqnarray}
where $\Z \equiv \sum_{\{\sigma\}} e^{-\beta \h(\sigma)}$ is the partition function (\ref{gib}) of the system described by the Ising Hamiltonian $\h(\sigma)$ of Eq.~(\ref{eq:IsingHam})
leading to 
\begin{eqnarray}
\Z &=& e^{N\beta J} [A_+^N + A_-^N ] , \\
A_{\pm} &=& \cosh(\beta h) \pm \sqrt{e^{-4\beta J} 	+ \sinh^2(\beta h )}. 
\label{zzz}
\end{eqnarray}
The last term  $\sum_{\{\sigma\}} e^{-\beta \tilde{\h}(\sigma,k)}$ in Eq.~(\ref{eq:P_f}) can be treated as a partition function of a system with Hamiltonian $\tilde{\h}$ defined via $-\beta \tilde{\h}(\sigma,k) = -\beta \h(\sigma) + i\lambda_k f(\sigma)$.

In equilibrium, the quantities of interest are related to the total spin magnetization (first moment) $m\in [-N,N]$ and correlations (second moment) $\eta\in [-N,N]$:
\begin{eqnarray}
\label{kk}
m={\sum}_{j=1}^N \sigma_j,\qquad \eta = {\sum}_{j=1}^N \sigma_j \sigma_{j+1}. 
\end{eqnarray}
From Eq.~(\ref{eq:P_f}) we then obtain the corresponding probability distributions 
\begin{subequations}
\label{eqs:Pmhdef}
\begin{eqnarray}
\label{eq:Pmdef}
P_m &=& \frac{1}{\Z(2N+1)}\sum_{k=-N}^{N} e^{-i\lambda_k m}\, {\rm Tr}[(V_k^{(1)})^N] , \\
P_{\eta} &=& \frac{1}{\Z(2N+1)}\sum_{k=-N}^{N} e^{-i\lambda_k \eta} \, {\rm Tr}[(V_k^{(2)})^N] , 
\label{eq:Phdef}
\end{eqnarray}
\end{subequations}
where the transfer matrices for $\tilde{\h}$'s are given by 
\begin{subequations}
\begin{eqnarray}
V_k^{(1)} &=& \begin{pmatrix}
e^{\beta J - \beta h - i\lambda_k} & e^{-\beta J}\\
e^{-\beta J} & e^{\beta J + \beta h + i\lambda_k}
\end{pmatrix} , \\
V_k^{(2)} &=& \begin{pmatrix}
e^{\beta J - \beta h + i\lambda_k} & e^{-\beta J - i\lambda_k}\\
e^{-\beta J - i\lambda_k} & e^{\beta J + \beta h + i\lambda_k}
\end{pmatrix} .
\end{eqnarray}
\end{subequations}
The final expressions for the probability distributions of total spin magnetization $m$ and correlation $\eta$ are 	
\begin{subequations}
\begin{eqnarray}
P_m & = & \frac{1}{\Z(2N+1)} \sum_{k=-N}^{N} e^{-i\lambda_km+N\beta J} \nonumber \\
& & \qquad \times [(A_{k,+}^{(1)})^N + (A_{k,-}^{(1)})^N] , \label{eq:P_m_anal} \\
A_{k,\pm}^{(1)} &=& \cosh(\beta h+i\lambda_k)  \pm \sqrt{e^{-4\beta J} + \sinh^2(\beta h+i\lambda_k)} , \nonumber \\
P_{\eta}&=& \frac{1}{\Z(2N+1)} \sum_{k=-N}^{N} e^{-i\lambda_k \eta + N \beta J+iN \lambda_k} \nonumber \\
& & \qquad \times [(A_{k,+}^{(2)})^N + (A_{k,-}^{(2)})^N] ,  \label{eq:P_h_anal} \\
A_{k,\pm}^{(2)} &=& \cosh(\beta h) - \sqrt{e^{-4\beta J-4i\lambda_k} + \sinh^2(\beta h)} \nonumber .
\end{eqnarray}
\end{subequations}
Using Eqs.~(\ref{zzz},\ref{kk})), the mean magnetization $\langle m\rangle$, its dispersion (variance) $\langle(\Delta m)^2\rangle$, and correlations $\langle\eta\rangle$ are given by
\begin{subequations}
\begin{eqnarray}
\label{ant}
&&\frac{\langle m\rangle}{N} =\frac{1}{N}\sum_{k=1}^N \langle \sigma_k\rangle=\frac{T}{N} \frac{\partial}{\partial  h}\ln\Z
=\frac{1}{N}\sum_{m=-N}^N m P_m\\
&& = \phi\sinh(\beta h)+{\cal O}\Big(\frac{A_-^N}{A_+^N}\Big), \quad \phi\equiv (\sinh^2(\beta h)+e^{-4\beta J})^{-1/2},\nonumber  \\
\label{gent}
&& \frac{\langle \eta\rangle}{N} =1-2e^{-4\beta J}\phi^2 (1+\phi\cosh(\beta h))^{-1}+{\cal O}\Big(\frac{A_-^N}{A_+^N}\Big),\\
&& \langle(\Delta m)^2\rangle \equiv \langle m^2\rangle-\langle m\rangle^2 =T^2\frac{\partial^2}{\partial h^2}\ln\Z.
\label{werpen}
\end{eqnarray}
\end{subequations}

\subsection{Zero magnetic field}
\label{subsec:zeromf}

\begin{figure}[t]
\includegraphics[clip,width=1\linewidth]{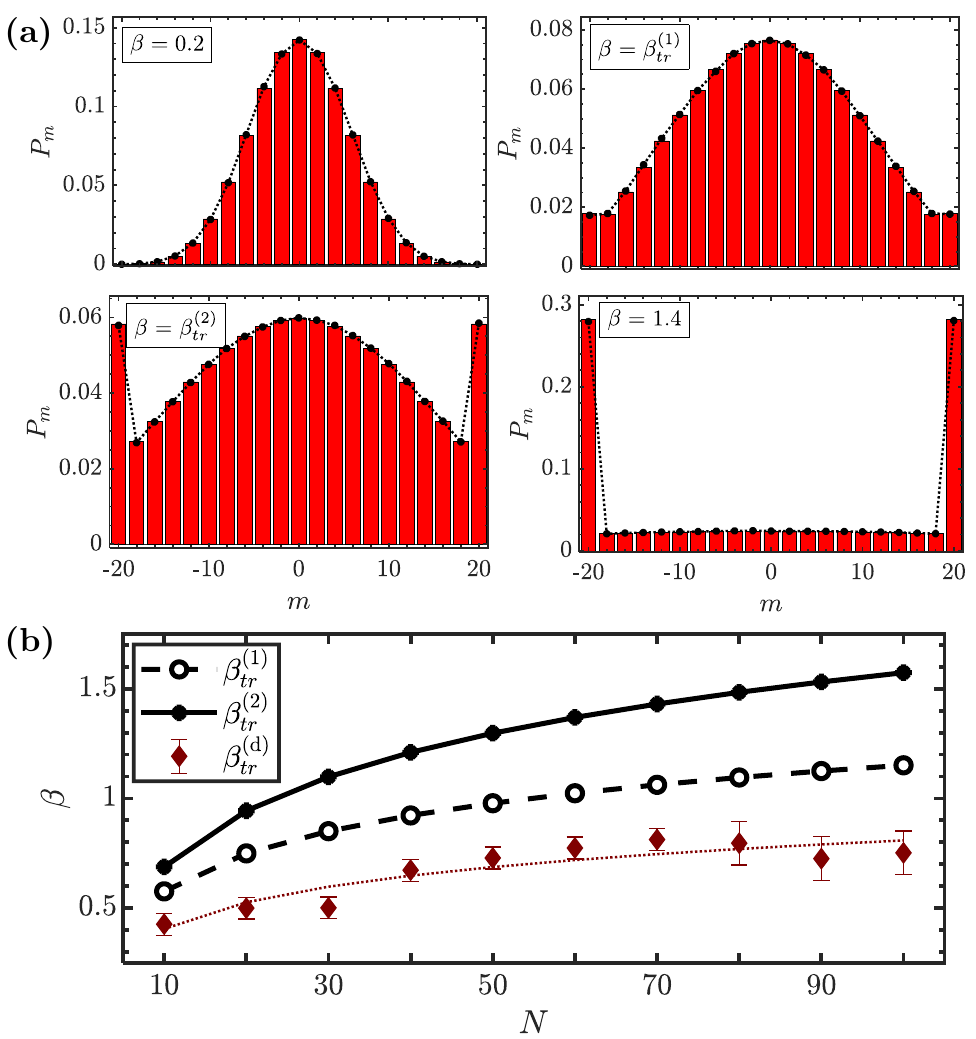}
\caption{
(a)~Probability distribution $P_m$ of magnetization $m$ for a chain of $N=20$ spins at different inverse temperatures $\beta=1/T$ (in units of $1/J$) and zero magnetic field $h=0$. Filled bars represent the analytical results of Eq.~(\ref{eq:P_m_anal}), while star markers connected with dotted lines correspond to the results of Monte Carlo simulations in  Sec.~\ref{sec:MCdyn}.
(b)~First and second inverse transition temperatures versus the number of spins $N$, $\beta_{tr}^{(1)} = 0.25 \ln(N)$ and $\beta_{tr}^{(2)} \approx 0.3864 \ln(N) - 0.21055$ (in units of $1/J$), as obtained from the $P_m$ distributions; 
and the dynamical inverse temperature $\beta_{tr}^{(\mathrm{d})}$ (maroon diamonds with error bars), as obtained from Monte Carlo simulations in Sec.~\ref{sec:MCdyn}, with the fit  $\beta_{tr}^{(\mathrm{d})} \approx 0.1755 \ln(N)$ (dotted line).}
\label{fig:Pm_beta}
\end{figure}
	
Consider first the case of $h=0$.
In Fig.~\ref{fig:Pm_beta}(a) we show the probability distribution $P_m$ of total magnetization $m$ of a chain of $N$ spins at different temperatures $T$.
Obviously, the mean magnetization vanishes at any temperature, $\langle m \rangle = \sum_m m P_m = 0$, while its variance is $(\Delta m)^2 \sim N$.
At high temperatures, $\beta = 1/T \ll 1/J$, the probability distribution of magnetization is peaked around $m=0$ corresponding to a paramagnetic chain with random orientation of spins. With decreasing the temperature (increasing $\beta$), the probability of paramagnetic states decreases while the probabilities of ferromagnetic states, $m=\pm N$, grow. 
We may define the first transition temperature $T_{tr}^{(1)}=1/\beta_{tr}^{(1)}$ at which the ferromagnetic state $|m|=N$ emerges with finite probability equal to that of the next least probable state $|m| = N-2$ with any one spin flipped, $P_{m=\pm N} = P_{m = \pm(N-2)}$, 
leading to $1 = Ne^{-4\beta J}$, and therefore
\begin{equation}
\label{t1}
\beta_{tr}^{(1)} = \frac{1}{4J} \ln(N).
\end{equation}
Equivalently, the entropy difference between these states is $\ln{N}$ while the energy difference is $4J$, and hence $T_{tr}^{(1)} \ln N = 4J$.
For still lower temperatures (larger $\beta$), the magnetization attains two dominant values $m=\pm N$ with all the spins aligned in the same direction due to the ferromagnetic interaction. We can then define the second transition temperature $T_{tr}^{(2)} = 1/\beta_{tr}^{(2)}$ at which the probabilities of ferromagnetic and paramagnetic states become equal, $P_{m= \pm N} = P_{m=0}$, obtaining, to a good approximation,
\begin{equation}
\label{t2}
\beta_{tr}^{(2)} \approx \frac{1}{5J} [ 2 \ln(N) - 1].
\end{equation}
The inverse transition temperatures for different lengths of the chain $N$ are shown in Fig.~\ref{fig:Pm_beta}(b). 
 
We emphasize that among the $2^N$ microscopic states, the two ferromagnetic configurations (degenerate for $h=0$) always have the largest probability as the minimal energy configurations as per Eq.~(\ref{gib}). Yet, at high temperatures, $T>T_{tr}^{(2)}$, the probability of  ferromagnetic state, corresponding to a single spin configuration, is smaller than the cumulative probability of the  paramagnetic state that includes many, $\binom{N}{N/2} \gg 1$, spin configurations with zero total magnetization.  

\begin{figure}[t]
\includegraphics[clip,width=1\linewidth]{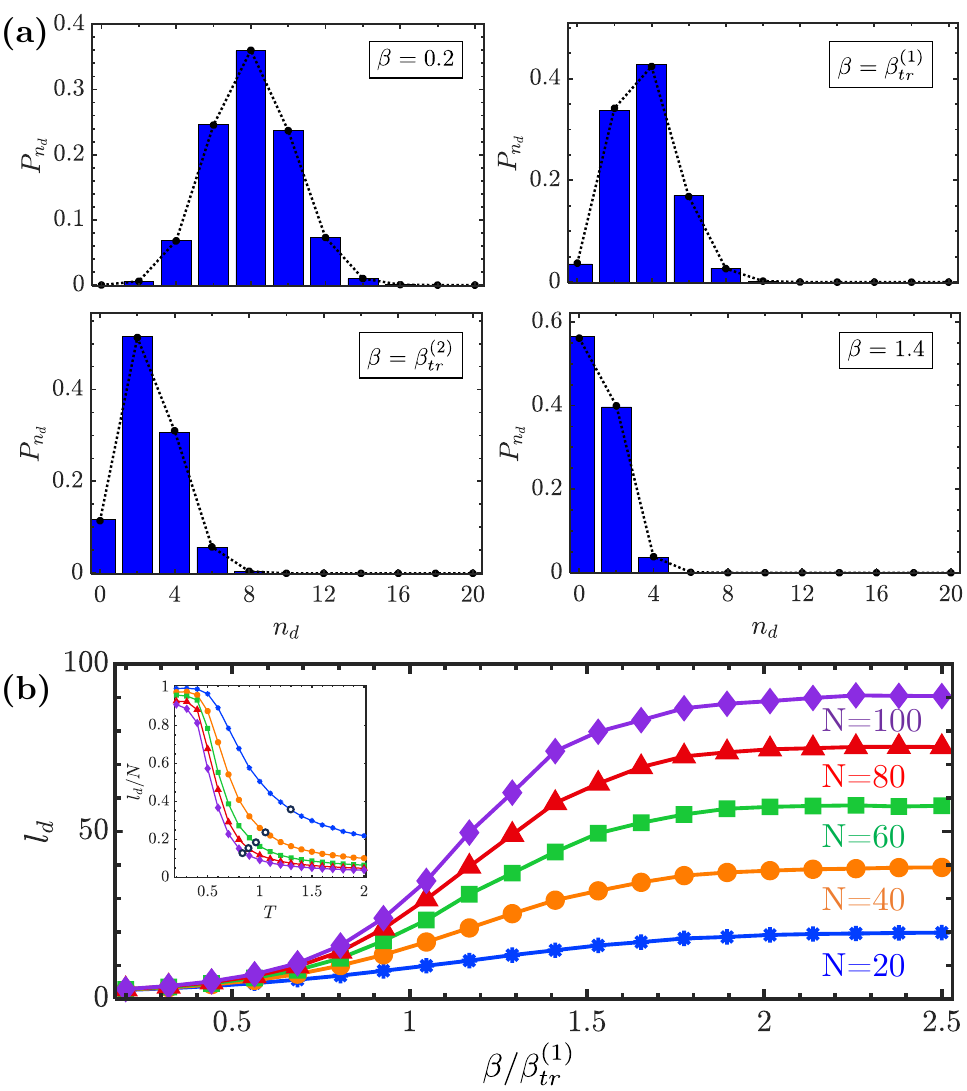}
\caption{
(a) Probability distribution $P_{n_d}$ of the number of domain walls $n_d$ for a chain of $N=20$ spins at different inverse temperatures $\beta=1/T$ (in units of $1/J$) and $h=0$. Filled bars represent analytical results, while star markers connected with dotted lines correspond to results of Monte Carlo simulations of Sec.~\ref{sec:MCdyn}.
(b)~Average domain length $l_d = \langle N/n_d\rangle$ of Eq.~(\ref{eq:ld}) versus the normalized inverse temperature $\beta/\beta_{tr}^{(1)}$ for chains of different lengths $N=20-100$, as obtained from the $P_{\eta}$ distribution.
Inset shows $l_d/N$ vs temperature $T$ with open circles denoting $T_{tr}^{(1)}$ for the corresponding $N$.
}
\label{fig:Pn_ld}
\end{figure}
	
Consider next the spin correlations. For a ferromagnetic chain with all the spins aligned, we have $\eta = N$, assuming periodic boundary conditions; cf.~(\ref{kk}).
For $n_d$ (even) domains with opposite spin orientations, separated by the same number of domain walls,  we have $\eta = N-2 n_d$, and therefore 
$n_d = (N-\eta)/2$.
Hence, the probability distribution of the number of domain walls and the average size $l_d$ of the domains are given by [cf.~(\ref{gent})]
\begin{align}
&P_{n_d} = P_{\eta = N-2 n_d}, \qquad
l_d =\Big\langle \frac{N}{n_d+\delta_{0\, n_d}}\Big\rangle.
\label{eq:ld}    
\end{align}

In Fig.~\ref{fig:Pn_ld}(a) we show the probability distribution of the number of domain walls, for the same parameters as in Fig.~\ref{fig:Pm_beta}(a). 
As expected, at low temperatures ($\beta > 1/J$) the most probable configurations are ferromagnetic, $n_d=0$, with $l_d \lesssim N$ as seen in Fig.~\ref{fig:Pn_ld}(b) where we show the average domain lengths for various temperatures and lengths of the chain. 
In the opposite limit of high temperatures ($\beta \ll 1/J$), there are many domains, $\langle n_d\rangle  \gg 1$, and $l_d \simeq N/\langle n_d \rangle \ll N$. 
Interestingly, at temperature $T_{tr}^{(2)}=1/\beta_{tr}^{(2)}$, the most probable number of domain walls is $n_d=2$ corresponding to two continuous ferromagnetic domains with opposite spin orientations, $l_d \lesssim N/2$. 
More precisely, we find that at temperatures $T_{tr}^{(1,2)} = 1/\beta_{tr}^{(1,2)}$, the average domain lengths $l_d^{(1,2)}$ of Eq.~(\ref{eq:ld}) grow with the chain size $N$ as  
\begin{equation}
\label{eq:ldpow}
l_d^{(1)} \propto N^{0.382} , \quad l_d^{(2)} \propto N^{0.757} .
\end{equation}
%
%
We verified that this power-law dependence of the domain sizes on the chain length is a specific feature of the equilibrium states at transition temperatures $T_{tr}^{(1,2)}$, while at other temperatures the power-law fits $l_d \propto N^{\gamma}$ have low fit quality for any $\gamma$.

\subsection{Finite magnetic field}

\begin{figure}[t]
\includegraphics[clip,width=0.75\linewidth]{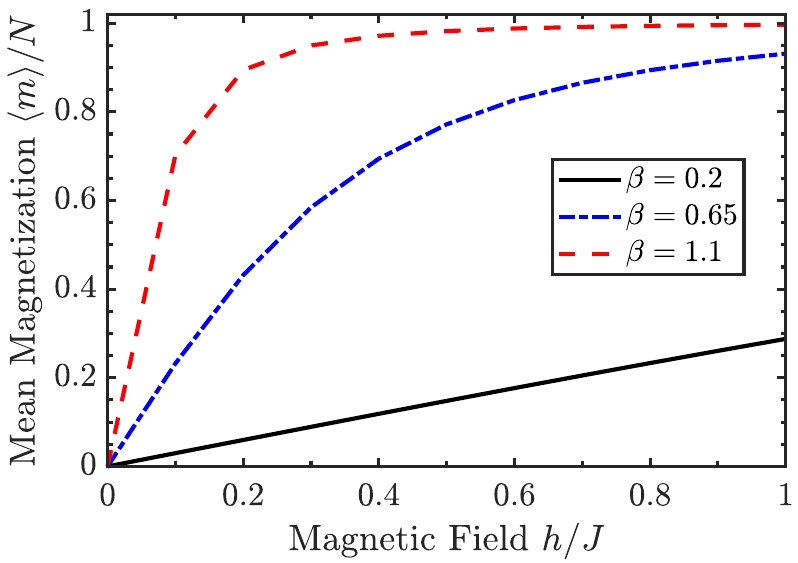}
\caption{Mean magnetization $\langle m \rangle$ vs magnetic field $h\geq 0$ at different inverse temperatures $\beta=1/T$ (in units of $1/J$) for a chain of $N=20$ spins.}
\label{fig:meanm}
\end{figure} 
	
Consider now the spin chain in the presence of magnetic field $h>0$. In Fig.~\ref{fig:meanm} we show the mean magnetization $\langle m \rangle$ of a spin chain at different temperatures $T$. We observe that, at high temperatures ($\beta \ll 1/J$), the magnetization grows slowly and linearly with the applied magnetic field, but at smaller temperatures ($\beta \gtrsim 1/J$), even a small external magnetic field $h \ll J$ can break the symmetry and strongly polarize the chain. 

\begin{figure}[t]
\includegraphics[clip,width=\linewidth]{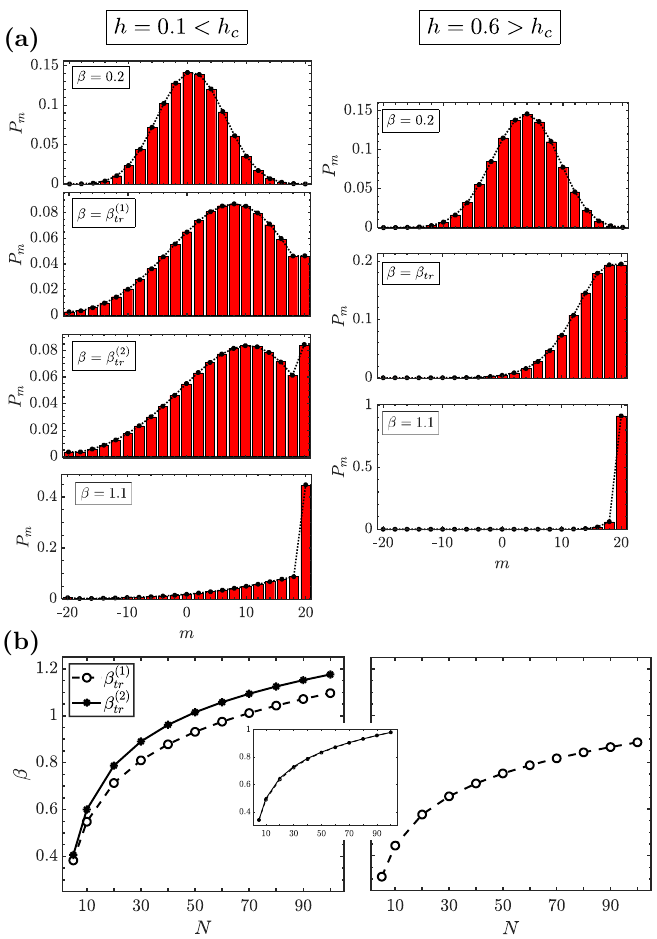}
\caption{(a) Probability distribution $P_m$ of magnetization $m$ for a chain of $N=20$ spins at different inverse temperatures $\beta=1/T$ and finite magnetic fields $h<h_c$ (left column) and $h>h_c$ (right column), with $h_c\simeq 0.45J$ as per Eq.~(\ref{eq:hc}).
(b)~The corresponding inverse transition temperatures (in units of $1/J$) versus the number of spins $N$, as obtained from the $P_m$ distributions. The inset shows the merging of $\beta_{tr}^{(1,2)}$ for $h = 0.35J \lessgtr h_c$ for $N \lessgtr 80$ respectively.}
\label{fig:Pmh}
\end{figure} 
	
Next, in Fig.~\ref{fig:Pmh}(a) we show the probability distributions of magnetization of a chain of $N$ spins for weak $h < h_{c}$ and strong $h> h_{c}$ magnetic fields, where $h_c$ is defined below in Eq.~(\ref{eq:hc}).  
At high temperatures, $\beta\ll 1/J$, the magnetic field shifts the paramagnetic peak of the probability distribution towards the higher values of the magnetization, $m>0$, while at small temperatures, $\beta \gtrsim 1/J$, only a single ferromagnetic configuration $m=N$ dominates. 
Similarly to the case of a zero magnetic field, we can define the (first) transition temperature $T_{tr}^{(1)}=1/\beta_{tr}^{(1)}$ as the temperature at which the probability of the ferromagnetic configuration becomes equal to the total probability of single spin-flip configurations, $P_{m=N} = P_{m= N-2}$, leading to $Ne^{-4\beta J - 2h} = 1$ and therefore 
\begin{equation}
\beta_{tr}^{(1)} = \frac{1}{4J+2h} \ln(N).  
\end{equation}
Note that in a sufficiently strong magnetic field $h \geq h_c$, $P_{m=N-2}$ is the largest probability other than that of the ferromagnetic state, $P_{m=N}$, and we can define only one transition temperature, i.e., the first and second transition temperatures coincide. But for weak magnetic fields, $h < h_c$, we can still define the second transition temperature $T_{tr}^{(2)}=1/\beta_{tr}^{(2)}$ at which the probability of the ferromagnetic state is equal to the peak probability of magnetization other than that corresponding to a single spin-flip, i.e., $P_{m=N} = \max[ P_{m< N-2} ]$. The critical magnetic field at which a peak of the probability distribution at $m \leq N-4$ still exists can be obtained from the conditions $P_{m=N} = P_{m=N-2} = P_{m= N-4}$ leading to 
\begin{equation}
h_{c} = 2J \, \frac{\ln(\frac{2N}{N+3})}{\ln(\frac{N+3}{2})} , \label{eq:hc}
\end{equation}
which can be approximated as $h_{c}/J \approx 2 \ln(2) / \ln(N/2)$ for $N \gg 1$. 
In Fig.~\ref{fig:Pmh}(b) we show the inverse transition temperatures for different lengths of the chain subject to weak $h<h_c$ and strong $h>h_c$ magnetic fields, where $h_c$ itself depends on the chain length $N$. Hence, if for some small $N$ and $h<h_c$ we have two transition temperatures, at sufficiently large $N \gtrsim 2^{2J/h+1}$, $\beta_{tr}^{(1)}$ and $\beta_{tr}^{(2)}$ can merge, see Fig.~\ref{fig:Pmh}(b) inset. 
	
\begin{figure}[t]
\includegraphics[clip,width=\linewidth]{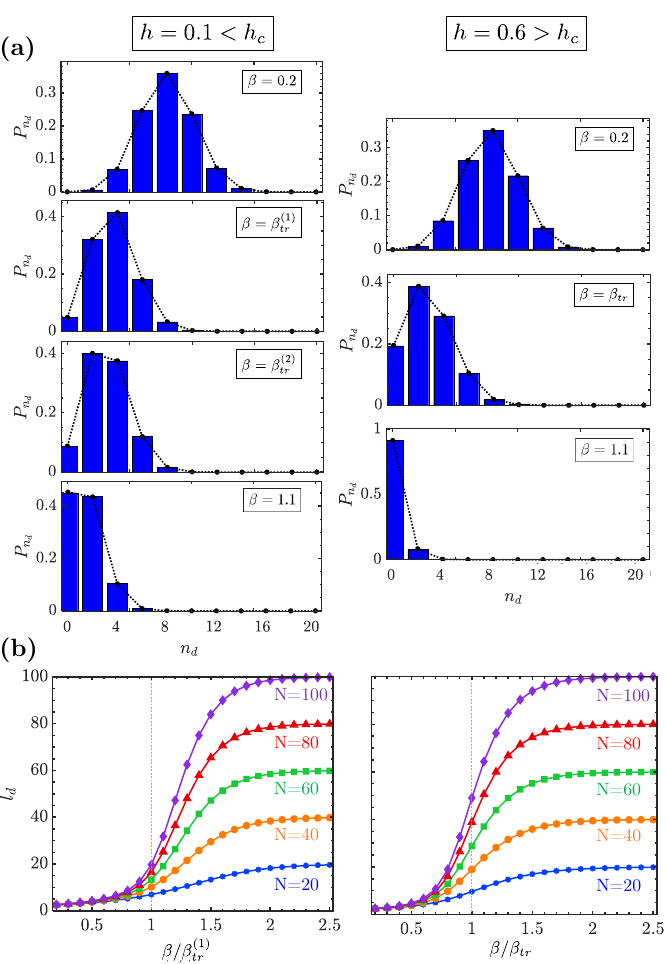}
\caption{(a) Probability distribution $P_{n_d}$ of the number of domain walls $n_d$ for a chain with $N=20$ spins at different inverse temperatures $\beta=1/T$ and magnetic fields $h<h_c$ (left column) and $h>h_c$ (right column).
(b)~The corresponding averaged domain lengths $l_d$ versus the normalized inverse temperature $\beta/\beta_{tr}^{(1)}$ for chains of different lengths $N=20-100$.}
\label{fig:Pndh}
\end{figure} 
    
In Fig.~\ref{fig:Pndh}(a) we show the probability distribution of the number of domain walls, for the same parameters as in   Fig.~\ref{fig:Pmh}(a). The variation of $P_{n_d}$ with the temperature is similar to that for $h=0$, but now the magnetic field $h \lesssim J$  polarizes the spins leading to a more homogeneous system attaining the ferromagnetic state at smaller $\beta \sim 1/J$ (larger $T$) with a single domain of length $l_d \lesssim N$, see Fig.~\ref{fig:Pndh}(b). 
Again, for $T = 1/\beta_{tr}^{(2)}$ (for $h<h_c$) or $T = 1/\beta_{tr}$  (for $h\geq h_c$) the most probable configurations correspond to $n_d=2$ domains with lengths $l_d \simeq N/2 \pm \langle m \rangle /2$ for the spin-up and spin-down domains. 
We find that, at temperature $T_{tr}^{(1)} = 1/\beta_{tr}^{(1)}$, the average domain length again follows the power law $l_d \propto N^{\gamma}$ with the exponents $\gamma \simeq 0.682, 1.029$ for $h=0.1, 0.6J$, respectively.


\section{Dynamics of 1d Ising chain}

\subsection{Dynamic Monte Carlo simulations}
\label{sec:MCdyn}

To better understand the results of the foregoing discussion and quantify equilibrium dynamics of the stochastic system, we perform numerical simulations of the dynamics of spin chains at different temperatures and magnetic fields. Direct simulations of rate equations~(\ref{eq:ddtpk}) are prohibitively difficult for $N \gtrsim 20$ spins as this would involve solving $2^N$ coupled differential equations for the probabilities of all the spin configurations. Instead, we use the standard algorithm \cite{Mitropolis1953,MC_Ising,MC_Gillespie,MC_Chemical} for Monte-Carlo simulations of the dynamics of the system governed by Hamiltonian (\ref{eq:IsingHam}) with the transition rates given by Eq.~(\ref{eq:Glauberrates}) \cite{Glauber_rates}. 

Briefly, starting with any microscopic state (spin configuration) 
$\{ \sigma \}_\mu \equiv \{ \sigma_1,\sigma_2, \ldots, \sigma_N \}$, we determine its total decay rate $\Gamma_\mu = \sum_{\nu\neq \mu} \Gamma_{\mu \nu}$ and set a waiting time $t_{\mu}$ 
chosen from a Poisson distribution with a mean $1/\Gamma_{\mu}$,
i.e.,  $t_{\mu} = -\ln(r)/\Gamma_{\mu}$, where $r\in [0,1]$ is random number from a uniform distribution.
At time $t_{\mu}$ we flip one spin determined according to the probabilities of individual spin-flips $\Gamma_{\mu \nu}/\Gamma_{\mu}$. 
We continue this process with the new spin configuration until the next spin-flip event, and so on, obtaining long-time trajectories of spin configurations. From many independent trajectories we can then determine ensemble-averaged quantities of interest, such as spin magnetization and correlations. 

In Figs. \ref{fig:Pm_beta}(a) and \ref{fig:Pmh}(a) we compare  equilibrium probability distributions of magnetization of the spin chain at different temperatures and external magnetic fields obtained from the analytic approach of the previous section and numerical Monte Carlo simulations. Similarly, in Figs.~\ref{fig:Pn_ld}(a) and \ref{fig:Pndh}(a) we compare the analytical and numerical results for the probability distributions of the number of domain walls. In all cases, we observe excellent agreement between the two methods. 

\begin{figure}[t]
\includegraphics[clip,width=0.8\linewidth]{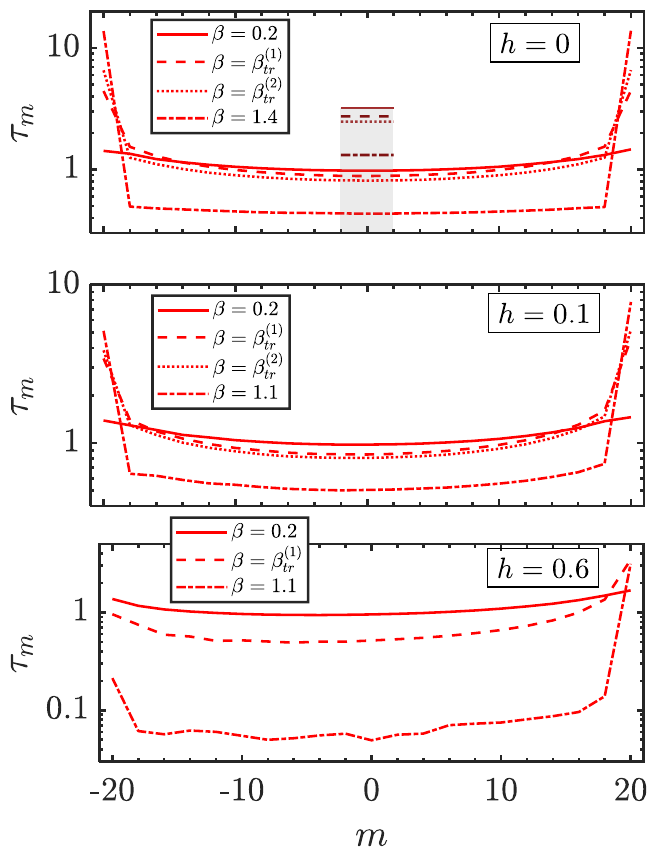}
\caption{
Residence times $\tau_m$ (in units of $1/\Gamma_0$) for the states with magnetization $m$ for a chain of $N=20$ spins at various magnetic fields $h=0,0.1,0.6J$ and different inverse temperatures $\beta=1/T$ (in units of $1/J$), as in Figs.~\ref{fig:Pm_beta}(a) (top panel) and \ref{fig:Pmh}(a) (middle and bottom panels). 
Maroon horizontal lines in the top panel for $h=0$ denote the residence times of the paramagnetic state with magnetization in the interval $|m| < \sqrt{N}/2$.}
\label{fig:tau_m}
\end{figure}

\subsection{Residence times of magnetization and dynamical transition temperature}

From the dynamical Monte Carlo simulations, we can also extract the residence times $\tau$ for various quantities of interest, i.e., the average duration of time intervals during which the corresponding quantity remains unchanged (or remains within some defined range). 
When calculating residence times $\tau_q$ of some quantity $q$, we select spin configurations $\{\sigma\}$ having prescribed values of that quantity and follow the time evolution of the system. Once the system attains a configuration with a different value of $q$ (or a value outside the defined range), we record the corresponding time interval $\Delta t$. This is then averaged over many $M\gg1$ Monte Carlo runs and over the Gibbs distribution $P(\{\sigma\}_j)$ of the initial configuration $\{\sigma\}_j$ with the set value of $q$: 
\begin{equation}
    \tau_q = \frac{1}{M}\sum_{i=1}^{M} \frac{\sum_{j} P(\{\sigma\}_j)\Delta t_{j}^{(i)}}{\sum_{j} P(\{\sigma\}_j)} .
\end{equation}
Both averaging procedures take place simultaneously if the configurations with different values of $q$ are selected from the long-time Monte Carlo trajectories.

In Fig.~\ref{fig:tau_m} we show the residence times $\tau_m$ of states with magnetization $m$ for different magnetic fields and temperatures, which should be compared and contrasted with Figs.~\ref{fig:Pm_beta}(a) and \ref{fig:Pmh}(a) for equilibrium probabilities $P_m$. 
At any temperature, the ferromagnetic state, $|m|=N$, has the longest residence time, while the paramagnetic state, $m=0$, has the shortest residence time for sufficiently small magnetic field.  
Stronger magnetic fields favor the collinearly oriented ferromagnet and reduce the lifetime of the ferromagnet with the opposite orientation of spins. For a given magnetic field, with increasing the temperature (decreasing $\beta$) the residence times of all the states approach  $\tau_m \to 1/\Gamma_0$ as expected from the transition rates of Eq.~(\ref{eq:Glauberrates}).

For simplicity, we focus on the case of zero magnetic field $h=0$. 
The long residence time $\tau_{m=\pm N}$ of the ferromagnetic state means that it is dynamically the most stable spin configuration having the smallest energy|and hence the largest equilibrium probability|at any temperature; see (\ref{gib}). This, however, does not mean that at any temperature the most probable magnetization should be $|m| = N$, cf. Figs.~\ref{fig:Pm_beta}(a) and \ref{fig:Pmh}(a), since at high temperatures the system can leave the ferromagnetic state and spend more time exploring the many spin configurations with magnetization around $m = 0$. 
Conversely, the short residence times of states with small magnetization is due to the rapid dynamics of the spin-flips of the higher energy configurations.
Hence, at high enough temperatures $T> T_{tr}^{(2)}$ the probability $P_m$ of magnetization $|m| \sim 0$ can be large due to the large number of contributing spin configurations visited often by the system. 

To rectify the apparent discrepancy between the dynamic and equilibrium quantities, $\tau_m$ and $P_m$, and compare fairly the lifetimes of ferromagnetic and paramagnetic states (as opposed to spin configurations), we define for the latter an interval of magnetizations around $m = 0$ with the width given by the standard deviation $\Delta m \sim \sqrt{N}$ that follows from Eq.~(\ref{werpen}) and Figs.~\ref{fig:Pm_beta}(a). In contrast, the ferromagnetic state is sharply peaked and therefore has vanishing dispersion.
The interval of values of $m$ that we assign to the paramagnetic state and the corresponding residence times are illustrated in the upper panel of Fig.~\ref{fig:tau_m}. The residence time 
of the paramagnetic state becomes smaller than that of the ferromagnetic state at a dynamical transition temperature $T^{(\mathrm{d})}_{tr}$ which is larger than the equilibrium transition temperatures, $T^{(\mathrm{d})}_{tr}>T^{(1)}_{tr}>T^{(2)}_{tr}$. 
In Fig.~\ref{fig:Pm_beta}(b) we plot the inverse temperature  $\beta^{(\mathrm{d})}_{tr} = 1/T^{(\mathrm{d})}_{tr}$ versus the chain length $N$, as obtained from the Monte Carlo simulations, and the fit 
\begin{equation}
\beta^{(\mathrm{d})}_{tr} \approx \frac{3}{17J} \ln(N). 
\end{equation}
For each value of $N$ we choose the interval $m \in [-\lfloor \sqrt{N} /2 \rceil, \lfloor \sqrt{N}/2 \rceil ]$ that is abruptly changing for certain values of $N$ (30 and 90) which explains the non-monotonic behaviour of $\beta^{(\mathrm{d})}_{tr}$ in Fig.~\ref{fig:Pm_beta}(b).

\subsection{Residence times of domain walls}

\begin{figure}[t]
\centering
\includegraphics[clip,width=0.8\linewidth]{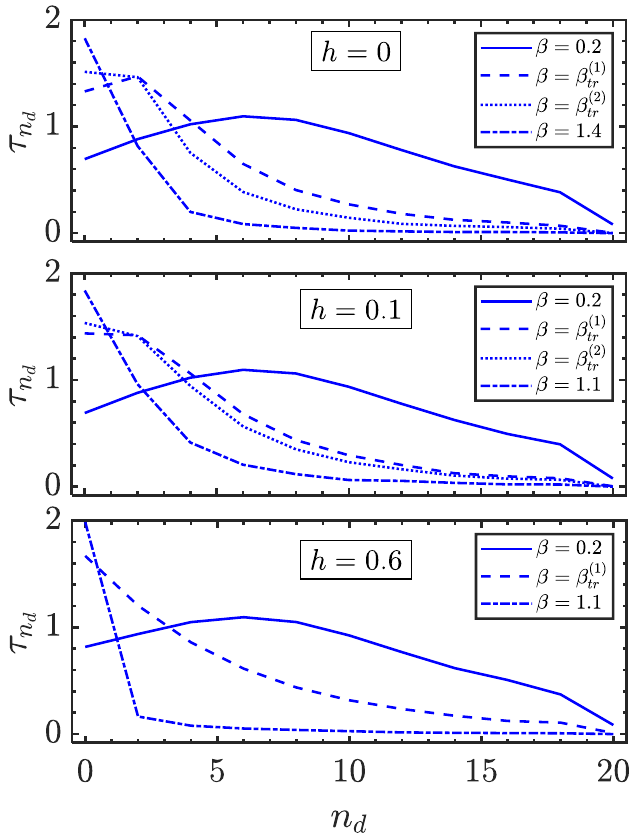}
\caption{
Residence times (in units of $1/\Gamma_0$) of the number of domain walls $n_d$ for a chain of $N=20$ spins at various magnetic fields $h=0,0.1,0.6J$ and different inverse temperatures $\beta=1/T$ (in units of $1/J$), as in Figs.~\ref{fig:Pn_ld}(a) (top panel) and \ref{fig:Pndh}(a) (middle and bottom panels).}
\label{fig:tau_nd}
\end{figure}

In Fig.~\ref{fig:tau_nd} we show the residence times $\tau_{n_d}$ for the numbers of domain walls $n_d$ in the system.
We observe that, overall, the residence times of $n_d$ at different magnetic fields and temperatures resemble the equilibrium probability distributions of the number of domain walls $P_{n_d}$, see Figs.~\ref{fig:Pn_ld}(a) and \ref{fig:Pndh}(a). 
Again, at small temperatures, $\beta \gtrsim 1$, the ferromagnetic configuration with $n_d=0$ has the longest residence time. With increasing the temperature (reducing $\beta$), the peak of $\tau_{n_d}$ shifts to the larger values of $n_d > 0$, but the positions of the peaks of $\tau_{n_d}$ and $P_{n_d}$ do not necessary coincide. The residence times $\tau_{n_d}$ are affected by two processes: Annihilation of two domain walls when they approach each other, the probability of which increases with $n_d$ since the distance between the domain walls $l_d \approx N/n_d$ decreases; and
creation of a pair of domain walls inside a continuous domain, the probability of which is larger for larger $l_d$ and thereby smaller $n_d$. The residence time is then peaked at a value of $n_d$ for which the total rate of these two processes is minimized.

A similar analysis of domain walls can lead to the notion of persistence in a 1D Ising-chain \cite{derrida, menon, review_persistence}. In an infinite system, the persistence shows the probability that a single spin has not flipped in time $t$ \cite{review_persistence}. 
For our Ising chain with $N$ spins, a more relevant definition of persistence would involve the fraction of spins $\rho(N)$ that do not flip at all times when the chain is subject to $T=0$ relaxation starting from a random initial state \cite{derrida}. The scaling relation $\rho(N)\sim N^{-2\theta}$ defines an independent dynamic persistence exponent $\theta>0$ \cite{derrida}. 
We also note that, in addition to the above two processes that determine the residence times of the domain walls, the persistence is also affected by the Brownian motion of the domain walls. 


\section{Structural transition in biopolymers}
\label{sec:biophys}

Many structural transitions in biopolymers can be modeled using the finite-size Ising chain with the parameters that depend on temperature. Examples include helix-coil transitions in DNA \cite{azbel}, formation of $\alpha$-helices from the coiled state in the protein secondary structure \cite{zimmo,go_go,book,qian,takano_eq,takano_neq}, and (de)naturation transition of the protein tertiary structure \cite{folding}. 
Below, we briefly review the application of the Ising model to the helix-coil transition and then show that our approach can predict the structure of intermediate states that are important for understanding protein functioning \cite{matt,ptit,privalov,finkel}.

The instantaneous configuration of the polymer is described by a set of helical and coiled regions.
To associate it phenomenologically with the Ising model \cite{book,qian}, we assign the spin variable
$\sigma_j=-1$ ($\sigma_j=1$) to the $j$'th helix (coil) region of the polymer, and assume the following free energy for the system
\begin{equation}
\label{eq:Fs}
F[\{\sigma\}]=-J{\sum}_{j=1}^{N} \sigma_i \sigma_{i+1} - h(T){\sum}_{j=1}^N \sigma_i,
\end{equation}
where $\{ \sigma \} \equiv \{ \sigma_1,\sigma_2, \ldots, \sigma_N \}$ is the configuration, and $N$ is the total number of regions \cite{badasyan}. 
In the polypeptide chain, the $\alpha$-helix is formed by hydrogen bonding between the monomers which are several units away from each other. Thus the formation of a bond facilitates bonding of its neighbors and leads to cooperativity described by $J>0$. Hydrophobic interactions and dipole-dipole forces also contribute to the cooperativity \cite{qian}. 

The parameter $h(T)$, playing the role of the effective magnetic field, favors helix formation at low temperatures, $h(T<T_0) < 0$, and coil formation at high temperatures, $h(T>T_0) > 0$, where $T_0\equiv 1/\beta_0$ is the helix-coil transition temperature at which $h(T_0)=0$. Experiments and {\it ab initio} calculations are consistent with a linear dependence of $h(T)$ on $T$ in the vicinity of $T_0$ \cite{book,qian,zimmo,go_go}:
\begin{equation}
h(T) = 2J (T/T_0 -1) = 2J (\beta_0/\beta -1) .    \label{eq:effIhT}
\end{equation}

We emphasize that the free energy for the coarse-grained helix and coil variables $\{\sigma_i\}$ in Eq.~(\ref{eq:Fs}) was obtained from a temperature-independent Hamiltonian of the full system upon integrating out all the other degrees of freedom of a biopolymer in solution, as is common in the Gibbs statistics \cite{armen_hc}. Yet $F[\{\sigma\}]$ in Eq.~(\ref{eq:Fs}) can be treated as the usual Hamiltonian of the Ising model for the spins $\{\sigma_i\}$ subject to the temperature dependent $h(T)$.
The effective magnetization $m < 0$ then corresponds to the helix dominated configurations, while $m > 0$ to the coil dominated configurations.
Starting from $T<T_0$, as we increase $T$, the mean magnetization changes from $\langle m\rangle/N \simeq -1$ to $\langle m\rangle/N \simeq 1$ for $T>T_0$; see (\ref{ant}, \ref{eq:effIhT}). In the vicinity of $\beta = \beta_0$, the equilibrium helix-coil transition is more abrupt (cooperative) for larger values of $J$ \cite{qian}, with the cooperativity characterized by $\frac{\langle n_d\rangle}{N}=(1+e^{2\beta_0 J})^{-1}$; see (\ref{gent}). In the highly-cooperative regime $\frac{\langle n_d\rangle}{N}\ll 1$, the equilibrium helix-coil transition resembles a real phase transition, which combines the features of first-order (discontinuous order parameter $\langle m\rangle/N$) and second-order (large correlation length) phase transitions \cite{book,qian}.

\begin{figure}[t]
\centering
\includegraphics[clip,width=1.0\linewidth]{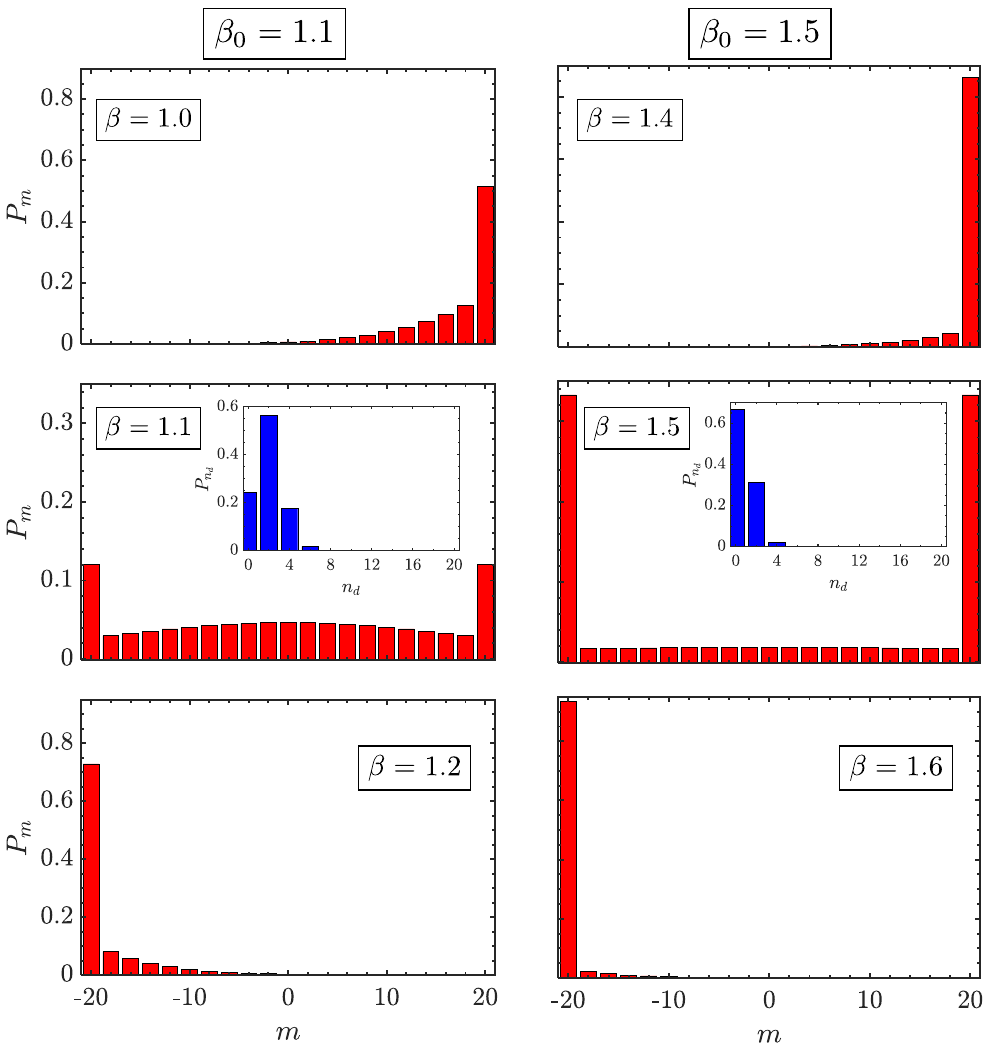}
\caption{Probability distribution $P_m$ of the effective magnetization of a chain of length $N=20$ with the inverse transition temperature $\beta_0=1.1$ (left column) and $\beta_0=1.5$ (right column) at three different temperatures $\beta <,=,> \beta_0$ (in units of $1/J$), for the model of Eqs.~(\ref{eq:Fs},\ref{eq:effIhT}).
The insets in the central panels show the corresponding probability distribution $P_{n_d}$ of the number of domain walls $n_d$ for $\beta=\beta_0$.}
\label{fig:PmFs}
\end{figure}

The normalized averages $\langle m\rangle/N$ and $\langle n_d\rangle/N$ do not reveal the structure of the intermediate state during the helix-coil thermal transition, but the probability distributions of Eqs.~(\ref{eqs:Pmhdef}) do contain such information. 
In Fig.~\ref{fig:PmFs} we show the probability distribution $P_m$ of the effective magnetization $m$ for two different values of the inverse transition temperature $\beta_0 \equiv 1/T_0$. 
At $T < T_0$ ($\beta > \beta_0$) and $h<0$, the most probable  magnetization $m=-N$ corresponds to a whole polymer in the helical state. Conversely, at $T > T_0$ ($\beta < \beta_0$) and $h>0$, the most probable magnetization $m=N$ corresponds to the uniformly coiled polymer. At the transition temperature $T=T_0$ ($\beta = \beta_0$) and $h=0$, we observe that, depending on the value of $\beta_0$, the intermediate values of the effective magnetization $|m| \sim 0$ have either appreciable or small probabilities (compare the left and right columns of Fig.~\ref{fig:PmFs}). Recalling the definition of the transition temperature from Sec.~\ref{sec:analytics}, we conclude that the former behavior is manifest when $\beta_0 \simeq \beta_{tr}^{(2)}$, while the latter behavior is manifest when $\beta_0 > \beta_{tr}^{(2)}$ and the helix-coil transition is a sharp transition between $m=-N$ and $m=N$ ``ferromagnetic'' states avoiding significantly populating mixed (``paramagnetic'') configurations with $|m| < N$.

In the insets of Fig.~\ref{fig:PmFs}, we show the probability distribution $P_{n_d}$ of the number of domain walls $n_d$ at transition temperature $\beta = \beta_0$. For $\beta_0 \simeq \beta_{tr}^{(2)}$ the most probable number of domain walls is $n_d =2$, while $n_d =4$ also has appreciable probability. This means that in the intermediate state, we have phase-separation whereby the helical and coiled regions coexist. 
But for $\beta_0 > \beta_{tr}^{(2)}$ the large probability $P_{n_d =0}$ of zero domain walls indicates that the coexistence of the helical and coiled phases is most probably absent. For larger $J$ the probabilities $P_{n_d >0}$ are further suppressed. Conversely, our simulations for longer chains with larger values of $\beta_{tr}^{(2)}$ lead to a probability distribution $P_{n_d}$ spread over a higher number of domain walls $n_d > 0$  indicating heterogeneous (effectively paramagnetic) intermediate state with many short helix and coil regions.%
\footnote{As an example, for $N=60$ and $J=1$ at the inverse transition temperature $\beta = \beta_0 = 1.1$ we obtain the probabilities 
$P_{n_d} = \{0.0037,0.0794,0.2686,0.3386,0.2124,0.0768,0.0175,0.0043, \ldots\}$ for $n_d=\{0,2,4,6,8,10,12,14,\ldots \}$ domain walls. Here $\beta_{tr}^{(2)}=1.37>\beta_0=1.1$. Taking the same parameters, but with $J=2$, we find for $\beta_{tr}^{(2)}=0.686<\beta_0=1.1$: $P_{n_d} = \{0.7824,0.2087,0.0086,0.0001, \ldots\}$.} 

There are thus three possibilities for the intermediate state at the transition temperature $T=T_0 \equiv 1/\beta_0$: 
For $\beta_0 > \beta_{tr}^{(2)}$ the intermediate state is nearly absent and the system abruptly transitions between the helical and coiled phases; for $\beta_0 \simeq \beta_{tr}^{(2)}$ the intermediate state is a phase-separated half-helical, half-coiled state; and for $\beta_0 < \beta_{tr}^{(2)}$ the intermediate state is heterogeneous, mixed helix-coil state. 

The Ising model is too simple to account for all the basic features of protein denaturation since, e.g., it neglects the fact that many proteins even in their fully denaturated state still contain certain permanent traces of the native state \cite{finkel}. Nevertheless, our conclusion that the structure of the intermediate state depends on two temperatures ($T_0$ and $T_{tr}^{(2)}$) are relevant whenever the intermediate state of a protein is important, e.g., for mapping protein folding pathways \cite{matt} and for understanding structurally disordered proteins with the intermediate state playing a functional role \cite{ptit,privalov,finkel}. In such states, up to $20-40\%$ of a protein can be disordered (i.e. coiled) \cite{finkel}.

\section{Summary and Discussion}
\label{sec:conclusions}


To summarize, our studies revealed a number of interesting properties of the finite-size $N$, ferromagnetic ($J>0$) Ising model at various temperatures $T$ and different external magnetic fields $h$. At high temperatures $T\gg J$ ($k_{\mathrm{B}} =1$) and small magnetic field, $h < J$, the most probable state is paramagnetic with random spin orientations. In the opposite limit of small temperatures, $T \ll J$, the most probable state is ferromagnetic, with all the spins having the same orientation, while even a small magnetic field $h\lesssim J$ can fully polarize the system. We identified equilibrium transition  temperatures $T^{(1,2)}_{{tr}}$ (functions $N$ and $h$) at which the ferromagnetic states already have appreciable probabilities. Interestingly, in the vicinity of transition temperature, even though the mean magnetization can be vanishing or small (for $h\neq 0$), the most probable microscopic configurations contain only a few domain walls, i.e., the spins tend to arrange into long domains with the same spin orientation. Another interesting result concerns the dynamics of the system at equilibrium: While the probability of ferromagnetic state is small at finite temperature and magnetic field, the residence times of the ferromagnetic configurations are always longer than those for all the other possible microscopic configurations. This means that once the system finds itself in a ferromagnetic configuration -- with however small probability -- it will remain there for a long time. When, however, we compare the ferromagnetic state to the paramagnetic state involving the set of spin configurations with nearly zero total magnetization, we can identify the dynamical transition temperature $T^{(\mathrm{d})}_{{tr}}$ for which the residence times of the ferromagnetic and paramagnetic states are equal, while for $T < T^{(\mathrm{d})}_{{tr}}$ the former lives longer than the later, and vice versa for  $T > T^{(\mathrm{d})}_{{tr}}$.

While the 1D Ising model does not exhibit a real phase transition in the thermodynamic limit, we note that analogs of temperatures $T^{(2)}_{tr}$ and $T^{(\mathrm{d)}}_{tr}$ exist for real phase transitions in systems with short-range interactions in more than one dimension or in systems with long-range interactions where the dimensionality is less important: The analog of $T^{(2)}_{tr}$ is the temperature of a real equilibrium first-order phase transition, while the analog of $T^{(\mathrm{d})}_{tr}$ is the temperature of dynamical ergodicity breaking, which in disordered systems (e.g. spin-glasses) is known to be higher than the temperature of static first-order phase transition \cite{luca,biro}.

We finally note that while we focused on the ferromagnetic Ising model $J>0$, the antiferromagnetic chain with $J<0$ and zero magnetic field $h=0$ is completely equivalent with the replacement of the magnetization and correlations with their staggered analogs: $m=\sum_j (-1)^j \sigma_j$ and $\eta = \sum_j (-1)^j \sigma_j \sigma_{j+1}$. For a finite field $h \neq 0$ the situation is, however, less straightforward, which will be the subject of future research.

\acknowledgments
This work was supported by the SCS of Armenia, grant No. 20TTAT-QTa003 (V.S., D.P., A.E.A.) and grant No. 22AA-1C023 (V.S.), and 
the EU QuantERA Project PACE-IN, GSRT grant No. T11EPA4-00015 and 
Horizon Europe programme HORIZON-CL4-2021-DIGITAL-EMERGING-01-30 via the project 101070144 EuRyQa (A.F.T. and D.P.).

\appendix

\section{Microcanonical distribution}
\label{microcano}

\begin{figure}[t]
\includegraphics[clip,width=\linewidth]{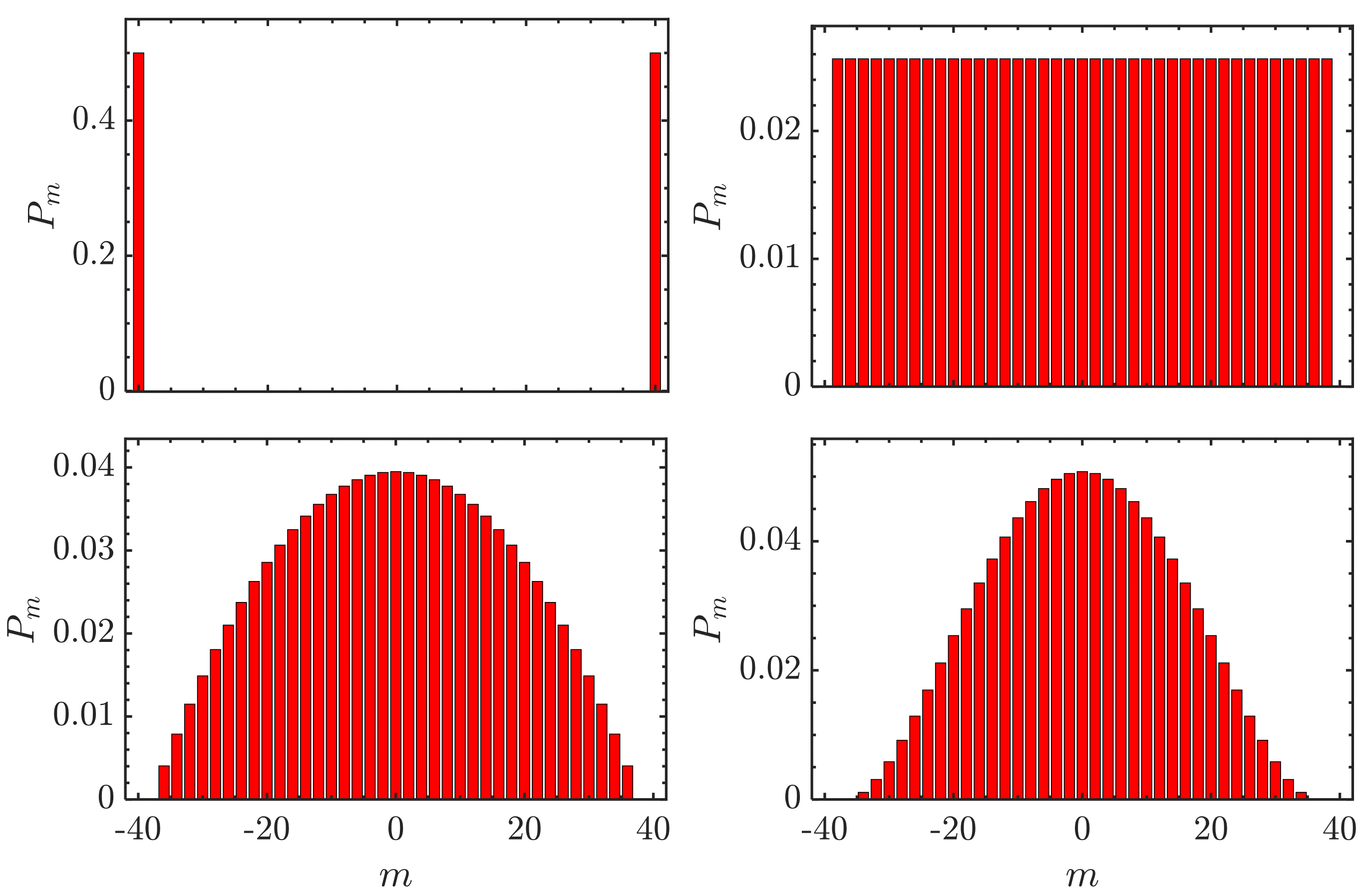}
\caption{Probability distributions $P_m$ of magnetisation for $N=40$ in a system with microcanonical distribution. Left to right, top to bottom are the cases with energies $E_0=-NJ$, $E_1=(-N+4)J$, $E_2=(-N+8)J$, $E_3=(-N+12)J$}
\label{fig:micro}
\end{figure}

The microcanonical distribution describes a physically isolated system at a fixed energy and provides an alternative approach to the macroscopic thermodynamics \cite{landau}. For macroscopic systems it is frequently equivalent to the canonical distribution \cite{landau}. The differences between these two distributions, however, are important and interesting for finite systems \cite{gross,berdi}. It was shown with several concrete examples (and led to a general belief) that the microcanonical distribution in finite systems facilitates the formation of various analogs of thermodynamic transitions \cite{gross,bachmann,hilbert}. Hence, the expectation would be that a microcanonic version of the one-dimensional Ising model would also demonstrate transitions in e.g. probability density of magnetization as a function of the energy. (Recall that the tuning parameter of the microcanonic distribution is the energy and not the temperature \cite{landau,berdi}). As we show below, this expectation is incorrect: transition effects found in the main text pertain to the canonical distribution, i.e. they do not exist in the microcanonical situation, where the natural control parameter is energy.

The energy spectrum of the system with Hamiltonian (\ref{eq:IsingHam}) (for simplicity we assume $h=0$) is discrete. In a microcanonical distribution, all spin configurations with the same energy $E$ have the same probability, while all configurations with a different energy have zero probability. Here $E$ is a control parameter of the distribution (recall that for the canonical distribution the control parameter is temperature). 

Obviously, the lowest energy states $E_0=-JN$ are ferromagnetic (zero domain walls) and we have only two configurations with magnetizations $m=N$ and $m=-N$ and equal probabilities $P_{m=\pm N} = 0.5$. The first excited state has energy $E_1=-JN+4J$ and $n_d=2$ domain walls. For each magnetization $m$, the lengths of the two domains are fixed since their sum is $N$ and diffrence is $m$. Hence $m$ assumes values $2-N \leq m \leq N-2$, and for each $m$ there is only one spin configuration. This means that probabilities for different magnetizations $m$ are the same, $P_m =1/(N-1)$, while $P_{m=+-N} = 0$.
For higher energy states, the probability distribution of magnetization is bell shaped and centered at $m=0$, see Fig.~\ref{fig:micro}. Thus transitions like the ones described in Figs.~\ref{fig:Pm_beta} do not occur upon increasing energy in the microcanonical distribution.


\begin{thebibliography}{100}

\bibitem{brush} S. G. Brush, History of the Lenz-Ising model, Rev. Mod. Phys. {\bf 39}, 883 (1967). 

\bibitem{lenz}W. Lenz, {\it Beitr\"age zum Verst\"andnis der magnetischen Eigenschaften in festen Kz\"orpern}, Physikalische Zeitschrift, {\bf 21}, 613 (1920).

\bibitem{ising}E. Ising, {\it Beitrag zur theorie des ferromagnetismus}, Zeit. Phys. {\bf 31}, 253 (1925).

\bibitem{baxter} R. J. Baxter, {\it Exactly Solved Models in Statistical Mechanics} (New York: Academic, 1982).

\bibitem{quasi1}
J.M. Bulled, M. Falsaperna, P.J. Saines, A.L. Goodwin, {\it Thermodynamic signatures of chain segmentation in dilute quasi-one dimensional Ising system}, arXiv:2212.06752. 

\bibitem{quasi2}
M. B. Yilmaz and F. M. Zimmermann, {\it Exact cluster size distribution in the one-dimensional Ising model}, Phys. Rev. E, {\bf 71}, 026127 (2005).

\bibitem{quasi3}
J.Kofinger and C. Dellago, Single-file water as a one-dimensional Ising model, New J. Phys. {\bf 12}, 09304 (2010).

\bibitem{azbel} M. Ya. Azbel, {\it Phase transitions in DNA}, 
Phys. Rev. A {\bf 20}, 1671 (1979).


\bibitem{zimmo}
B.H. Zimm {\it et al.}, PNAS, {\bf 45}, 1601 (1970).

\bibitem{go_go}
M. Go {\it et al.}, J. Chem. Phys. {\bf 52}, 2060 (1970).

\bibitem{book} 
K. Sneppen and G. Zocchi, {\it Physics in Molecular Biology}
(Cambridge University Press, Cambridge, 2005).

\bibitem{qian}
A.J. Doig, {\it Recent advances in helix–coil theory}, Biophys. Chem. {\bf 101-102}, 281 (2002).


\bibitem{folding} A. Bakk and J.S. Hoye, {\it One-dimensional Ising model applied to protein folding}, Physica A, {\bf 323} 504-18 (2003).

\bibitem{armen_hc}
A. E. Allahverdyan, S. G. Gevorkian, and A. L. Simonian, {\it Kinetics of helix-coil transition},
Europhysics Lett. (EPL) {\bf 86}, 53002 (2009).

\bibitem{armen1}
A.E. Allahverdyan and A. Galstyan, {\it On the Maximum a Posteriori Estimation of Hidden Markov Processes}, UAI 2009: Proceedings of the Twenty-Fifth Conference on Uncertainty in Artificial Intelligence, 1–9, 2009; arXiv:0906.1980.

\bibitem{armen2}
A.E. Allahverdyan and A. Galstyan, {\it Active inference for binary symmetric hidden Markov models}, J Stat. Phys., {\bf 161}, 452-466 (2015).

\bibitem{bruce} A.D. Bruce, {\it Probability density functions for collective coordinates in Ising-like systems}, J. Phys. C \textbf{14}, 3667 (1981).

\bibitem{racz}
T. Antal, M. Droz, and Z. Racz, {\it Probability distribution of magnetization in the one-dimensional Ising model: effects of boundary conditions}, J. Phys. A {\bf 37}, 1465 (2004).


\bibitem{campo} Z.Xu and A. Del Campo, {\it Probing the full distribution of many-body observables by single-qubit interferometry}, Physical Review Letters, {\bf 122}, 160602 (2019).
 
\bibitem{Glauber_rates} R.J. Glauber, 
{\it Time-dependent statistics of the Ising model}, 
J. Math. Phys. \textbf{4}, 294-307 (1963).
		
\bibitem{Mitropolis1953}
N. Metropolis, A.W. Rosenbluth, M.N. Rosenbluth, and A.H. Teller, 
{\it Equation of State Calculations by Fast Computing Machines}, 
J. Chem. Phys. \textbf{21}, 1087 (1953).
	
\bibitem{MC_Ising}
A. B. Bortz, M. H. Kalos, and J. L. Lebowitz, 
{\it A new algorithm for Monte Carlo simulation of Ising spin systems},
J. Comput. Phys. \textbf{17}, 10 (1975).

\bibitem{MC_Gillespie}
D. T. Gillespie, 
{\it A General Method for Numerically Simulating the Stochastic Time Evolution of Coupled Chemical Reactions}, 
J. Comput. Phys. \textbf{22}, 403 (1976).

\bibitem{MC_Chemical}
K. A. Fichthorn and W. H. Weinberg, 
{\it Theoretical foundations of dynamical Monte Carlo simulations}, 
J. Chem. Phys. \textbf{95}, 1090-1096 (1991).
	





\bibitem{Allahverdyan2000}
A.E. Allahverdyan, Th.M. Nieuwenhuizen,
{\it Steady adiabatic state: Its thermodynamics, entropy production, energy dissipation and violation of Onsager relations},
Phys. Rev. E {\bf 62}, 845, (2000).

\bibitem{derrida} B. Derrida, V. Hakim, and V. Pasquier, {\it Exact First-Passage Exponents of 1D Domain Growth: Relation to a Reaction-Diffusion Model}, Phys. Rev. Lett. {\bf 75}, 751 (1995). 

\bibitem{menon} G. I. Menon, P Ray, P Shukla {\it Persistence in one-dimensional Ising models with parallel dynamics}, Phys. Rev. E {\bf 64}, 046102 (2001). 

\bibitem{review_persistence}S. N. Majumdar, {\it Persistence in nonequilibrium systems},
Current Science, {\bf 77}, 370-375 (1999). 

\bibitem{takano_eq}
M. Takano, K. Nagayama, and A. Suyama, {\it Investigating a link between all-atom model simulation and the Ising-based theory on the helix–coil transition: Equilibrium statistical mechanics}, J. Chem. Phys. {\bf 116}, 2219-2228 (2002).

\bibitem{takano_neq}
M. Takano, H.K. Nakamura, K. Nagayama, and A. Suyama, {\it Investigating a link between all-atom model simulation and the Ising-based theory on the helix–coil transition. II. Nonstationary properties}, 
J. Chem. Phys. {\bf 118}, 10312-22 (2003).

\bibitem{matt}C.R. Matthews, {\it Pathway of protein folding}, Ann. Rev. Biochem. {\bf 62}, 653-683 (1993).


\bibitem{ptit}
O.B. Ptitsyn, {\it Molten globule and protein folding}, Adv. Protein Chem. {\bf 47}, 83-229 (1995).

\bibitem{privalov}P.L. Privalov, {\it Intermediate states in protein folding}, J. Molecular Biology {\bf 258}, 707-725 (1996). 

\bibitem{finkel} V.N. Uversky, A.V. Finkelstein, {\it Life in phases: Intra-and inter-molecular phase transitions in protein solutions}, Biomolecules {\bf 9}, 842 (2019).

\bibitem{badasyan} A. V. Badasyan, A. Giacometti, Y. Sh. Mamasakhlisov, V. F. Morozov, and A. S. Benight, {\it Microscopic formulation of the Zimm-Bragg model for the helix-coil transition}, Phys. Rev. E {\bf 81} 021921 (2010)

\bibitem{luca}L. Leuzzi and T.M. Nieuwenhuizen, {\it Thermodynamics of the glassy state} (CRC press, 2007).

\bibitem{biro} L. Berthier and G. Biroli, {\it Theoretical perspective on the glass transition and amorphous materials}, Rev. Mod. Phys. {\bf 83}, 587 (2011).

\bibitem{landau}  L.D. Landau and E.M. Lifshitz, {\it Statistical Physics: Volume 5} (Elsevier, 2013).

\bibitem{gross}
D. H. E. Gross, {\it Microcanonical Thermodynamics} (World Scientific, Singapore, 2001).

\bibitem{berdi} 
V.L. Berdichevsky, {\it Thermodynamics of Chaos and Order} (Addison Wesley Longman, Essex, England, 1997).

\bibitem{bachmann}
S. Schnabel, D. T. Seaton, D. P. Landau, and M. Bachmann, {\it Microcanonical entropy inflection points: Key to systematic understanding of transitions in finite systems}, Phys. Rev. E {\bf 84}, 011127 (2011). 

\bibitem{hilbert}
J. Dunkel and S. Hilbert, {\it Phase transitions in small systems: Microcanonical vs. canonical ensembles}, Physica A, {\bf 370}, 390-406 (2006).








\end{thebibliography}
\end{document}